\documentclass[12pt,preprint]{aastex}
\shorttitle{LH41-1042}

\shortauthors{Neugent et al.}

\begin{document}

\title{The Discovery of a Rare WO-type Wolf-Rayet Star\\ in the Large Magellanic Cloud\altaffilmark{1}}

\author{Kathryn F.\ Neugent and Philip Massey}
\affil{Lowell Observatory, 1400 W Mars Hill Road, Flagstaff, AZ 86001;\\ kneugent@lowell.edu; phil.massey@lowell.edu}

\author{Nidia Morrell}
\affil{Las Campanas Observatory, Carnegie Observatories, Casilla 601, La Serena, Chile; nmorrell@lco.cl}

\altaffiltext{1}{This paper includes data gathered with the 6.5 meter Magellan Telescope located at Las Campanas Observatory, Chile.}

\begin{abstract}
While observing OB stars within the most crowded regions of the Large Magellanic Cloud, we happened upon a new Wolf-Rayet star in Lucke-Hodge 41, the rich OB association that contains S Doradus and numerous other massive stars. At first glance the spectrum resembled that of a WC4 star, but closer examination showed strong OVI $\lambda \lambda$ 3811, 34 lines, leading us to classify it as a WO4. This is only the second known WO in the LMC, and the first known WO4 (the other being a WO3). This rarity is to be expected due to these stars' short lifespans as they represent the most advanced evolutionary stage in a massive star's lifetime before exploding as SNe. This discovery shows that while the majority of WRs within the LMC have been discovered, there may be a few WRs left to be found.
\end{abstract}


\section{Wolf-Rayet Stars in the Large Magellanic Cloud}
\begin{figure}
\plotone{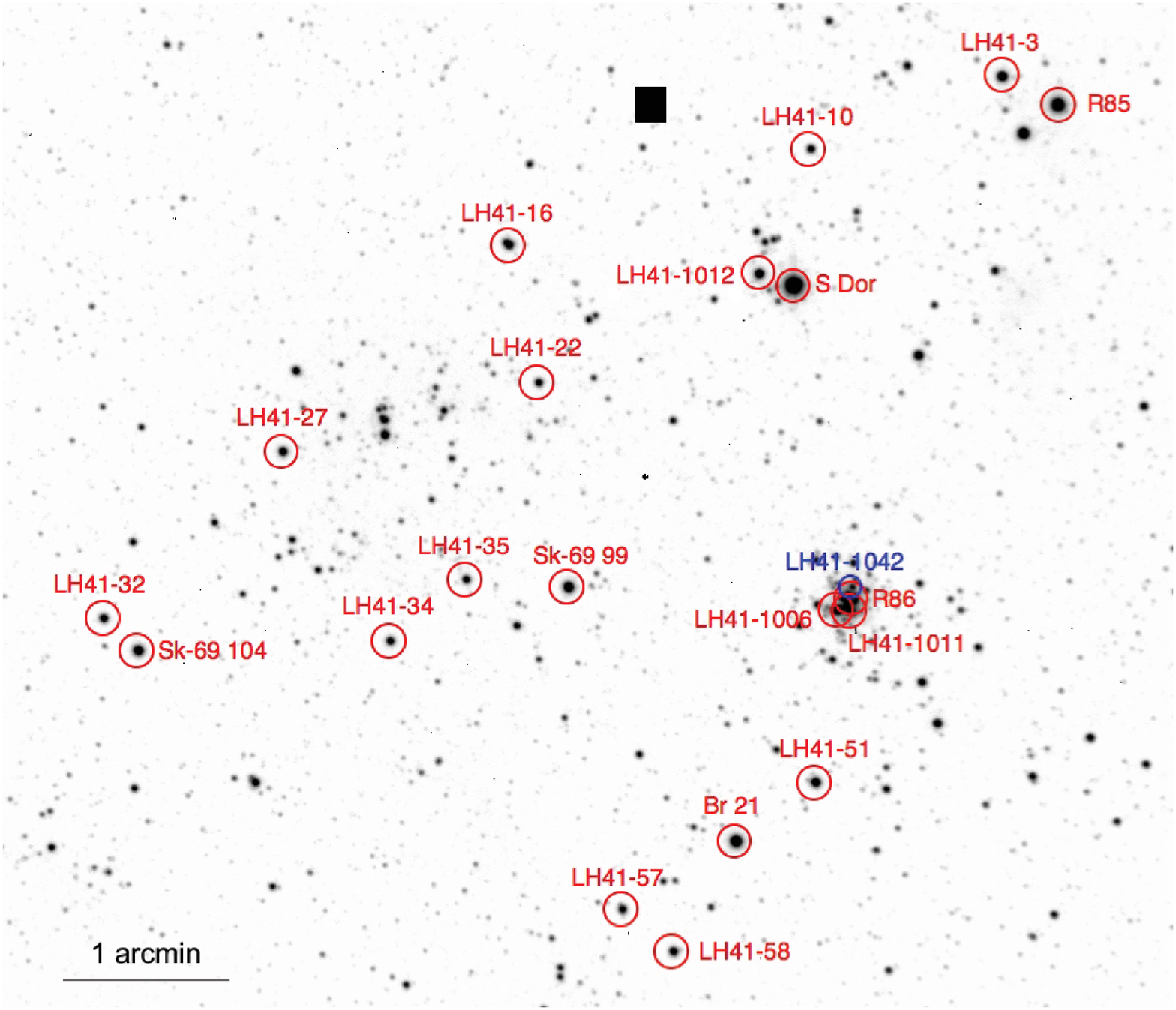}
\caption{\label{fig:locations} The locations of all known supergiants and O stars in LH-41. The stars' spectral types and $V$ magnitudes are shown in Table 1 and come from Massey et al.\ 2000. Our newly found WR star is labeled LH41-1042 and denoted by a blue circle and label. Massey et al.\ 2000 list an additional O8 V star in LH41 but we don't include it here because of an ambiguity in its name and location.}
\end{figure}

On the northern edge of the Large Magellanic Cloud's (LMC) central bar lies one of the galaxy's most visually striking OB associations, Lucke-Hodge 41 (LH41, also known as NGC 1910; see Lucke \& Hodge 1970), possibly second only to 30 Doradus in its massive star population. This young association contains a rich collection of stars including two of the seven known luminous blue variables (LBVs) in the LMC (one being the well-known S Doradus), a Wolf-Rayet (WR) star / B supergiant pair, two A-type supergiants, as well as a yellow supergiant, just to name a few (Massey et al.\ 2000). It was in this exciting region (see Figure~\ref{fig:locations} and Table 1) that the authors accidentally discovered a new WR star (LH41-1042) while observing crowded OB stars. This new WR star has strong enough oxygen lines to be classified as a WO, making it the second known WO star in the LMC and the first known WO4 (the other being a WO3). This discovery is both exciting and surprising given the long history of LMC WR surveys. However, as Figure 1 shows, given its location in such a densely crowded region near R86, LH41-1006 and LH41-1011, the fact that it hasn't been discovered until now is understandable. 

Westerlund \& Rogers (1959) published the first systematic search for WR stars in the LMC after using slitless spectroscopy (objective prism) to discover 50 WRs. Two decades later, Azzopardi \& Breysacher (1979, 1980) completed an even more powerful objective prism survey using an interference filter to further reduce the effects of crowding. This increased the number of known WRs in the the LMC to 100. Accurate spectral types of these 100 LMC WRs were published by Breysacher (1981). In that paper, Breysacher (1981) estimated that $44 \pm 20$ LMC WRs were left to be discovered. Breysacher (1981) further hypothesized that the majority of these undiscovered WRs would be found deep within the cores of dense H II regions where slitless spectroscopy often fails. Indeed, the most recent catalogue of LMC WRs (Breysacher et al.\ 1999; BAT99) lists 134 separate WRs, well within the initial estimate of $144 \pm 20$. 

As stated in Breysacher et al.\ (1999), the most recently discovered WRs are indeed located in crowded H II regions, much like the WR we discuss here. However, unlike our newly-found WR of WO-type, the majority (11 out of 12) of the new stars presented in BAT99 are of WN-type. This is to be expected since, as discussed in Neugent \& Massey (2011), the strongest emission feature in WCs is nearly 4$\times$ stronger than the strongest line in WNs (Conti \& Massey 1989), making WNs much more difficult to detect. Therefore before BAT99, the ratio of un-detected WRs in the LMC was previously skewed towards WNs. However, as we have proven, the occasional WR not of WN-type may yet be discovered.

\section{Our New WO-type Wolf-Rayet Star}
We came across the newly found WO4 star while characterizing the massive star content of Lucke-Hodge 41.  The results of the broader investigation will be reported elsewhere; here we discuss the new WR. We used the Magellan Echellette (MagE) on the 6.5-m Clay telescope. The spectrum has a resolution of 4100, and covers the entire optical range from 3100\AA-1$\mu$m. The exposure was 600 sec in total, observed in three 200 sec segments, while the star was dithered along the 1-arcsec width slit. The data were reduced as described in Massey et al.\ (2012), \S2.2.

\begin{figure}
\epsscale{0.70}
\plotone{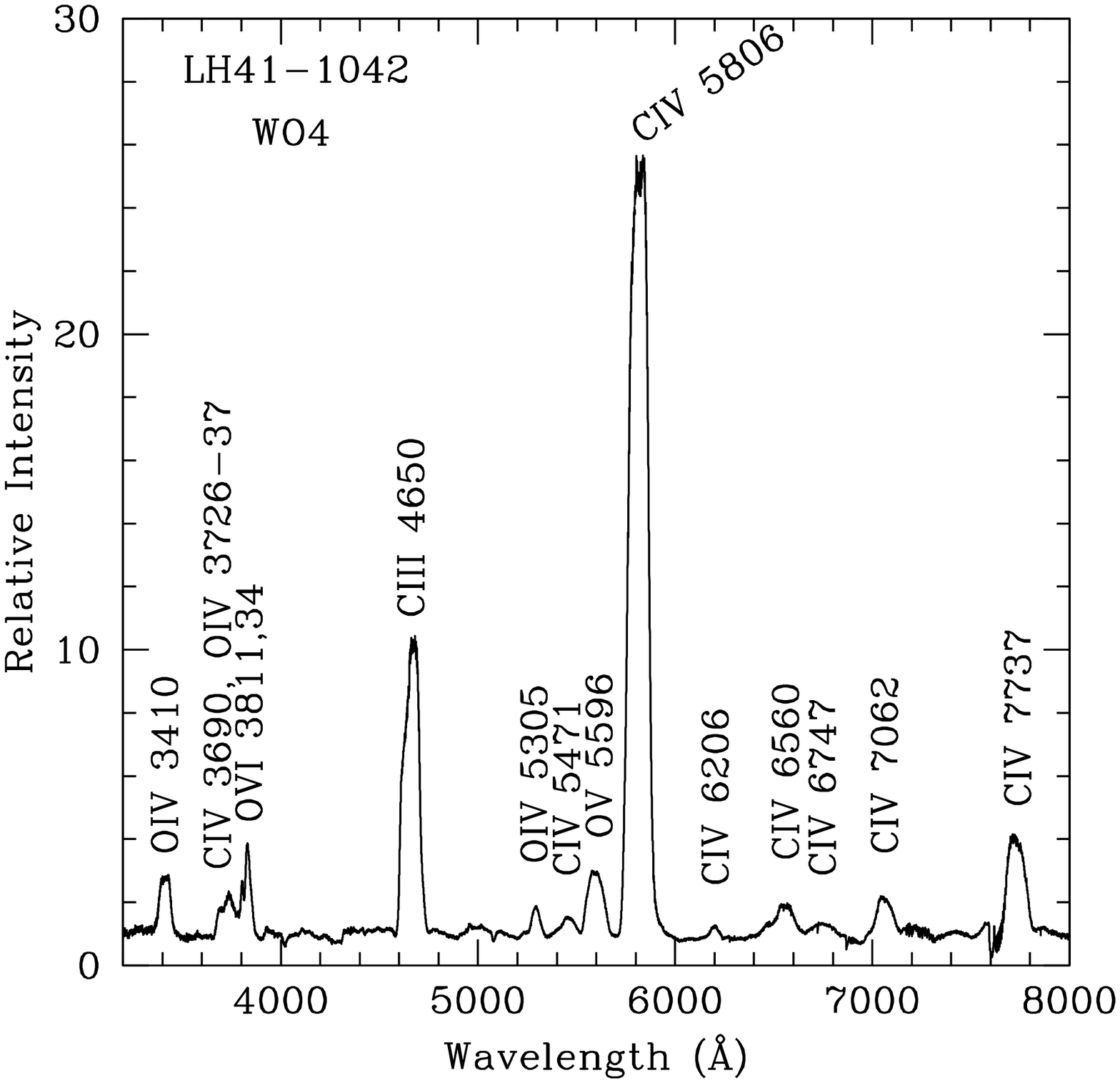}
\epsscale{0.33}
\plotone{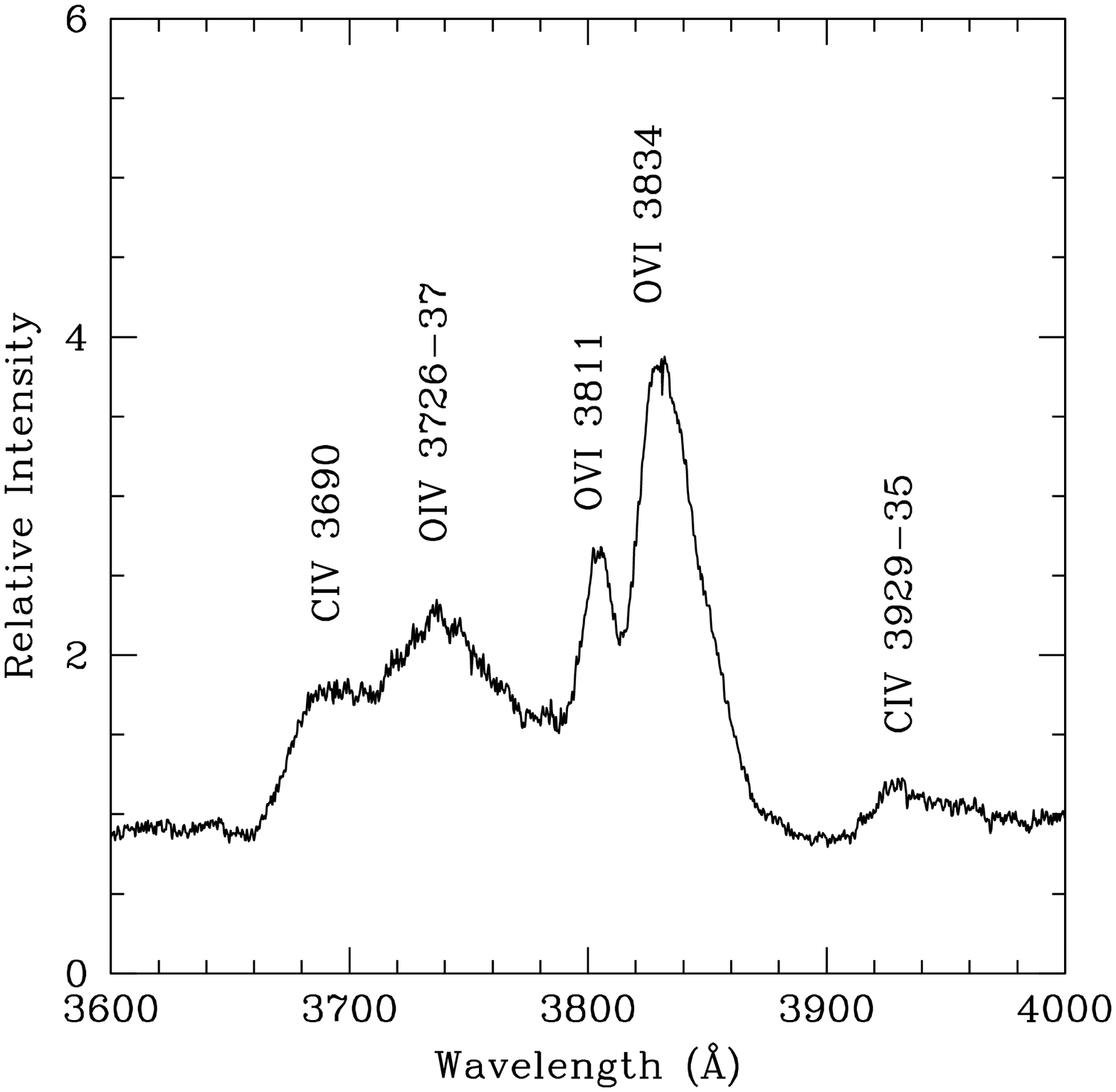}
\plotone{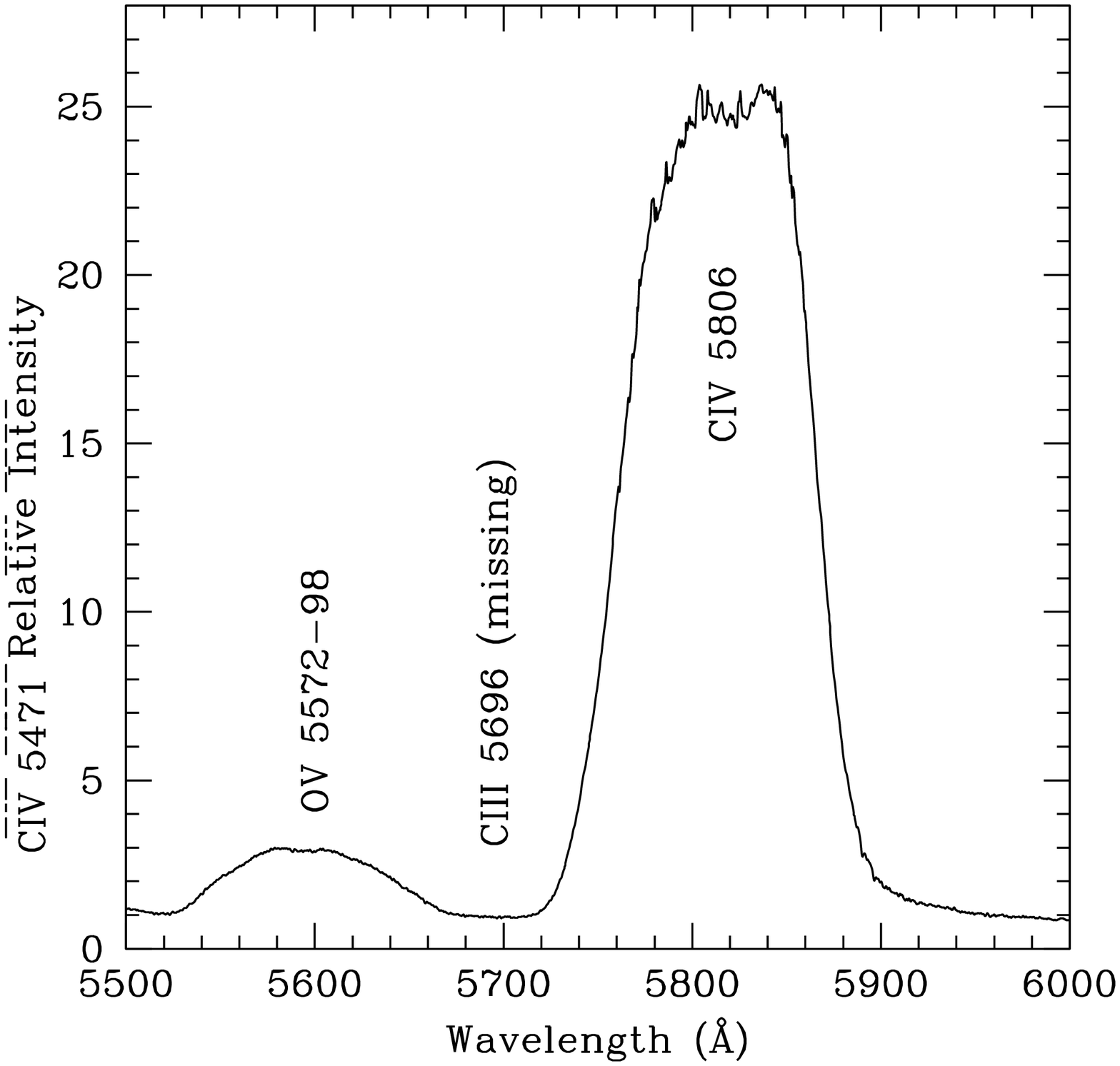}
\caption{\label{fig:spectra} Main spectral lines in LH41-1042. Note the OVI $\lambda\lambda 3811,34$ lines which make this star a WO.}
\end{figure}

Our new WR star is located at 05:18:10.91 -69:13:11.5 (J2000) and is listed in Simbad as [L72] LH 41-1042. This designation comes from Massey et al.\ (2000) who extended the numbering scheme of stars in the Lucke-Hodge 41 association (Lucke 1972) by denoting new additions as 1XXX. For LH41-1042, Massey et al.\ (2000) measured $V = 13.95$, $B-V = 0.31$ and $U-B = -1.38$. In Figure~\ref{fig:spectra}, we show the main spectral lines in LH41-1042 which we identified using Torres \& Massey (1987). The lack of the CIII $\lambda 5696$ line, and the moderate strength of the OV $\lambda 5572-98$ complex would lead to a WC4 classification, consistent with the broadness of the lines ($\sim$ 90\AA). However, we identify strong OVI $\lambda\lambda 3811,34$ lines, characteristic of WO-type stars (Barlow \& Hummer 1982). Based upon the equivalent widths (EWs) of the OVI lines relative to the OV and CIV $\lambda 5806$ lines, we classify this as a WO4, following Crowther et al.\ (1998); in their Figure 7, the star would fall near MS4, with a $\log$ EW($\frac{\rm OVI}{\rm CIV}$)$=-1.3$ and a $\log$ EW($\frac{\rm OVI}{\rm OV}$)$=-0.15$. The EW values are shown in Table~\ref{tab:WOew}.

\section{Discussion}
BAT99 lists 134 WRs in the LMC. Out of these 134 stars, only one of them is listed as a WO (BAT99-123\footnote{This star is also known as Sand 2 and was analyzed by Crowther et al.\ (2000) as well as Kinsburgh et al.\ (1995). These analyses played an important role in our current understanding of massive WO-type stars.}, a WO3). This rarity is not unexpected; out of the 154 WRs in M31 and the 206 WRs in M33, none are WOs (Neugent et al.\ 2012 and Neugent \& Massey 2011). Though rare, these stars aren't especially unusual; in essence, they are WC stars with abnormally strong OVI $\lambda \lambda$ 3811, 34 lines (Barlow \& Hummer 1982). Thus, the spectrum of our star, a WO4, is almost identical to that of a WC4, which is the most common type of WC in the LMC. The strong OVI $\lambda \lambda$ 3811, 34 lines that characterize a WO spectrum are due to high ionization, although such stars are also very rich in carbon and oxygen, as deduced by recombination analyses (Kingsburgh et al.\ 1995), and supported by non-LTE models (Crowther et al.\ 2000). Therefore, they most likely represent the most advanced (and short-lived) evolutionary stage in the life of a massive star before they explode as SNe. Current evolutionary models have a hard time producing WO stars (taken to have a number abundance $\frac{\rm C+O}{\rm He} > 1$ at the surface); see, for example, Table 1 in Georgy et al.\ (2012).

In addition to our new WO4 star, several other WRs have been discovered since the publication of BAT99 and these new stars are listed in Table~\ref{tab:newWRs}. As expected, the majority (in this case, all) of the newly found WRs are of WN-type and many are located in crowded regions. Two WRs have also been ``demoted" to Of-stars since the catalogue. These are also listed in Table~\ref{tab:newWRs}. This brings the total number of WRs in the LMC to 138 stars (134 stars from BAT99 + 6 newly discovered stars - 2 demoted stars = 138 stars). 

Reid \& Parker (2012) mention finding several additional WRs in the LMC using spectroscopic followup to an H$\alpha$ imaging survey, but they plan to give details (coordinates, spectral types) in a future paper. W.\ Reid (2012, private communication) has confirmed that our newly found WO4 star is not among them. H$\alpha$ imaging should be effective at finding very late-type WNs (WN9-11 stars, sometimes called Ofpe/WN9), but will be less effective for earlier WNs or WCs and WOs (see the survey of H$\alpha$ emission stars in other Local Group galaxies by Massey et al.\ 2007).

Thanks to the work of Breysacher and Azzopardi, along with many others, the majority of WRs within the LMC have already been discovered. However, as our discovery shows, especially in the most crowded regions of the LMC, there are still a few stray WRs left to be found. The continuing discoveries of WRs, especially of WO-type, are important because of the stars' scant numbers and poorly-understood properties. With each new discovery comes an opportunity to better understand these short-lived and thus extremely rare stars. 

\acknowledgements
We would like to thank Brian Skiff for doing a literature search on recently-found WRs as well as the anonymous referee their thorough reading of the manuscript. Additionally, we are especially grateful for the (as always) excellent support we received at Las Campanas during our observing run. This work was supported by the National Science Foundation under AST-1008020.

\begin{deluxetable}{l l r l}
\tablecaption{\label{tab:CoolStars} Supergiants and O Stars in LH-41\tablenotemark{a}}
\tablewidth{0pt}
\tablehead{
\colhead{Star}
&\colhead{$V$}
&\colhead{$B-V$}
&\colhead{Type}}
\startdata
S Dor & 9.32 & 0.11 & LBV \\
R85 & 10.53 & 0.16 & LBV (A Ie) \\
Br 21 & 11.28 & -0.07 & B1 Ia + WN3 \\
R86 & 11.52 & -0.15 & B0.2 I \\
LH 41-1006 & 11.78 & -0.10 & B0.5 I \\
Sk -69$^\circ$ 99 & 11.80 & 0.06 & A0 I \\
Sk -69$^\circ$ 104 & 12.09 & -0.20 & O7 III(f) \\
LH 41-3 & 12.10 & 0.06 & A2 I \\
LH 41-51 & 12.33 & -0.16 & O9.5 I \\
LH 41-1011 & 12.35	 & -0.17 & B0.2 I \\
LH 41-1012 & 12.42 & -0.16 & O9.5 I \\
LH 41-27 & 12.79 & -0.13 & O7.5 If \\
LH 41-16 & 12.93 & -0.16 & O8.5 III(f) \\
LH 41-32 & 13.03 & -0.20 & O4 III \\
LH 41-58 & 13.15 & -0.14 & O8.5 III \\
LH 41-34 & 13.15 & -0.16 & O6 III(f) \\
LH 41-22 & 13.38 & 0.67 & F5 I \\
LH 41-35 & 13.39 & -0.21 & O7 III(f) \\
LH 41-57 & 13.53 & -0.15 & O9.5 V \\
LH 41-10 & 13.54 & -0.20 & O8.5 V \\
\enddata
\tablenotetext{a}{From Massey et al.\ 2000.}
\end{deluxetable}

\begin{deluxetable}{l l}
\tablecaption{\label{tab:WOew} Equivalent Widths}
\tablewidth{0pt}
\tablehead{
\colhead{Line}
&\colhead{EW (\AA)}}
\startdata
OVI 3811-34 & 145 \\
OV 5596 & 200\\
CIV 5806 & 2800\\
\enddata
\end{deluxetable}

\begin{deluxetable}{l l l l r l r}
\tablecaption{\label{tab:newWRs} WRs Discovered Or Demoted Post-BAT 99}
\tablewidth{0pt}
\tablehead{
\colhead{Star}
&\colhead{$\alpha$(J2000)}
&\colhead{$\delta$(J2000)}
&\colhead{$V$}
&\colhead{$B-V$}
&\colhead{Type}
&\colhead{Ref\tablenotemark{a}}}
\startdata
[M2002] 15666 & 04 53 03.78 & -69 23 51.7 & 14.39 & -0.18 & WN3b + abs & 1 \\
BAT99-5a & 04 55 07.60 & -69 12 31.7 & 15.11 & -0.12 & WN3 & 2 \\
BAT99-15a & 05 02 58.24 & -69 14 02.3 & 15.19 & -0.15 & WN3h+abs & 2 \\
Sk -69$^\circ$ 194 & 05 34 36.08 & -69 45 36.5 & 11.91 & -0.07 & B0I + WN & 3 \\
LH 90$\beta$-6 & 05 35 58.72 & -69 11 52.3 & 12.94 & 0.13 & B I + WN & 3 \\
P93-1732 & 05 38 55.51 & -69 04 26.8 & 16.08 & 0.58 & WN5 & 4 \\
Br 58 & 05 35 42.27 & -69 11 53.9 & 14.13 & 0.49 & O3If*/WN6 & 3 \\
BAT99-093 & 05 37 51.35 & -69 09 46.8 & 13.54 & -0.08 & O3If* & 4 \\
\enddata
\tablenotetext{a}{References: 1 = Zaritsky et al.\ 2004; 2 = Howarth \& Walborn 2012; 3 = Massey et al.\ 2000; 4 = Evans et al.\ 2011}
\end{deluxetable}

\end{document}